\documentclass[preprint,12pt]{elsarticle}

\usepackage{amssymb}
\usepackage{subeqnarray}
\usepackage{cases}

\journal{Physics Letters A}

\begin{document}

\begin{frontmatter}



\title{Vortex-lattice structures in rotating Bose-Fermi superfluid mixtures}

\author[label1]{Wen Wen}
\affiliation[label1]{organization={College of Science, Hohai University},
               city={Nanjing 210098},
               country={China}}

\author[label2]{Lu Zhou}
\affiliation[label2]{organization={Department of Physics, School of Physics and Electronic Science, East China Normal University},
               city={Shanghai 200241},
              country={China}}

\author[label1]{Zhenjun Zhang}

\author[label3]{ Hui-jun Li}
\affiliation[label3]{organization={Institute of Nonlinear Physics and Department of Physics, Zhejiang Normal University},
               city={Jinhua  321004},
               country={China}}

\begin{abstract}

The system of Bose-Fermi superfluid mixture offers a playground to explore rich macroscopic quantum phenomena.
In a recent experiment of Yao {\it et al.} [Phys. Rev. Lett. {\bf 117}, 145301 (2016)], $^{41}$K-$^{6}$Li
superfluid mixture is implemented. Coupled quantized vortices are generated via rotating the superfluid mixture,
and a few unconventional behaviors on the formations of vortex numbers are observed, which can be traced to boson-fermion interactions.
Here we provide a theoretical insight into the unconventional behaviors observed in the experiment. To this end, the orbital-free density functional theory is hired, and its utility is validated by making comparison of the numerical results and a microscopic theory for vortex lattices in strongly interacting Fermi superfluids alone. We also predict interesting phenomena which can be readily explored experimentally, including
the novel structures of bosonic-fermionic vortices, and attractive interactions between the bosonic and fermionic vortices.

\end{abstract}




\end{frontmatter}








\section{Introduction}\label{secI}

One of the characteristic properties of superfluids is that they respond to rotation by forming
quantized vortices \cite{pet2008}. Systems of ultracold quantum gases stand out as they provide an ideal platform
to study quantized vortices with precise controls and broad tunability \cite{fet2001,fet2009}. Since the experimental realization
of Bose-Einstein condensation (BEC) in atomic clouds, triangular vortex lattice being
a conventional structure has been created in rotating Bose superfluid \cite{abo2001, mad2000},
and widely investigated theoretically \cite{fed2001, mak2003, lobo2004}.

Like bosons, ultracold fermioins also offer insights into macroscopic quantum phenomena.
A tunable Feshbach resonance provides a unique opportunity to access
pairing and superfluidity in the crossover between a Bardeen-Cooper-Schrieffer (BCS) state of
largely overlapping pairs of fermions and a BEC state of fermion dimers \cite{blo2009,zai2009}.
On the cusp of this BCS-BEC crossover, there exists a strongly interacting regime-the so-called unitary limit \cite{zwe2012},
which becomes the subject of numerous experiments \cite{raj2023,pac2022,yan2022,xiang2021,fang2021,die2021}. Convincing proof
of superfluidity along the BCS-BEC crossover is obtained through the observation of quantized vortices \cite{zwi2005,sch2007}.
In contrast to weakly interacting Bose superfluids well understood in terms of the Gross-Pitaevskii (GP) theory \cite{fed2001, mak2003, lobo2004}, interacting fermions requires a complete microscopic approach which is more complicated and computationally intensive \cite{gio2008,abul2012}.
To account for Pauli exclusion, one should resort to the orbital-based density functional theory (DFT), such as
Bogoliubov-de Gennes (BdG) equations \cite{dav2004,sim2015,hu2007,kon2021,fan2022} or superfluid local
density approximation (SLDA) \cite{aur2003,abu2011,bul2014,hos2022,kop2021,bar2023}. Both of the two methods require to solve a huge system of non-linear
equations in a self-consistent way \cite{dav2004,sim2015,kop2021,aur2003,hu2007,kon2021,fan2022,abu2011,bul2014,hos2022,bar2023}.
To reveal the properties and configurations of vortex lattices under experimental conditions, a spatial coarse graining of the BdG equations was performed \cite{sim2015}. Later on, an asymmetric SLDA was applied to study the formations of vortex lattices in spin-imbalanced unitary Fermi gases harmonically confined in two-dimensional (2D) traps \cite{kop2021}.

While the long-sought goal of simultaneous superfluidity in mixtures of $^{4}$He-$^{3}$He
still remains elusive due to strong interisotope interactions, Bose-Fermi superfluidity
has been realized in atomic gas mixtures of $^{7}$Li-$^{6}$Li \cite{fer2014, tak2016}, $^{41}$K-$^{6}$Li \cite{yao2016} and $^{174}$Yb-$^{6}$Li \cite{roy2017}. In the recent experiment of $^{41}$K-$^{6}$Li atoms \cite{yao2016}, the conclusive evidence of
the double superfluidity is first provided by producing coupled vortex lattices.
By carrying out a series of vortex-number measurements,
the significant effects of boson-fermion interactions on the formation and decay of the coupled vortices are studied.
However, there is no theoretical interpretation thus far.

One can envisage that the vortex lattices in Bose-Fermi superfluid mixtures can possess novel properties
compared to a single superfluid. Apart from the intercomponent
interactions as in two-component BECs, the distinct quantum statistical properties of the two atom species
give rise to far richer vortex-lattice structures. The intercomponent
interaction drives the two-component BECs \cite{mat1999,and2000,sch2004,kas2005}
going through transition from miscible to immiscible phases \cite{pu1997}.
For equal masses and equal intracomponent interactions, vortex lattices in the miscible phase have triangular,
rectangular, square, and double-core structures \cite{muel2002,kas2003}, while the lattices in the immiscible phase
are featured by stripes and interwoven vortex sheets \cite{kas2003,kas2009,ken2018,mas2011}.
Asymmetric systems where the two components have different masses and different intracomponent interactions,
support coreless vortex lattices in the miscible phase, and rotating droplets and giant skyrmions in the
immiscible phase \cite{mas2011,aft2012, ska2020}. In addition, unequal masses can result in  homogeneous infinite vortex lattices
having some notable geometries \cite{min2018,min2019}. These studies have been focused on
the somewhat less computationally intensive 2D realm.

It is thus natural to question what new features come about \cite{ling2014,jiang2017,pan2017,ogr2020}, if we couple a rotating, weakly
interacting Bose superfluid and a rotating, strongly interacting Fermi superfluid as in the experiment.
Within the lowest Landau level approximation \cite{jiang2017}, the structures of vortex lattices in the BCS-BEC crossover is
determined by minimizing the total free energy of Bose-Fermi mixtures. The transition of the vortex-core structure of the fermionic component
on the BCS side as a function of repulsive Bose-Fermi interactions is investigated by solving coupled BdG and GP equations self-consistently \cite{pan2017}.

Our work aims at studying vortex lattices in a experimentally relevant trapped rotating Bose-Fermi mixture \cite{yao2016} in the unitary limit, and unravelling the structures
of the coupled vortices through a miscible-immiscible transition. To this end, our approach relies on the so-called orbital-free DFT \cite{kim2004,sal2008,lsal2008,
adh2008,wen2008}, which for fermions is not written in terms of single-particle orbitals as the orbital-based DFT
aforementioned, but only
in terms of a single macroscopic wavefunction \cite{wen2008}.
The orbital-free DFT provides an attractive computationally practical method for simulating
huge systems under realistic conditions. In the framework of the orbital-free DFT,
the rotational properties of Bose-Fermi superfluid mixtures confined in a 2D harmonic trap
\cite{ling2014} and in a tight toroidal trap \cite{ogr2020} have been recently investigated.

We first apply the orbital-free DFT to study vortex lattices in strongly interacting Fermi superfluids alone.
The obtained vortex numbers and critical frequencies in the BCS-BEC crossover
are consistent with those from the BdG equations \cite{sim2015}.
Subsequently, we carry out extensive numerical simulations on the structures of vortex lattices as functions of repulsive boson-fermion interactions
and rotation frequencies. It is found that the effects of the repulsive interspecies interaction
on the vortex lattices of the Bose and Fermi superfluids are very different,
which may provide a theoretical insight into the experiment. Various structures of bosonic-fermionic vortices
in phase-separated states under various rotation frequencies are further revealed. This study is of particular interest under the experimentally accessible conditions as it can present a comparison between the experiment and the numerical results of the simplified DFT, as well as illustrating
unique features of Bose-Fermi superfluid mixture.

The paper is organized as follows. In Sec.\;\ref{secII}, we present the theoretical formalism
for a rotating, harmonically trapped Bose-Fermi superfluid mixture, and the numerical methods.
First in Sec.\;\ref{secIII}, we study the properties of vortex lattices in strongly
interacting Fermi superfluids alone numerically and analytically. Next in Sec.\;\ref{secIV}, we study the effects of repulsive
boson-fermion interactions on the vortex-lattice structures.
Finally, we conclude in Sec.\;\ref{secV} with a summary of our results and
an outlook to future research.

\section{theoretical model and numerical approach}{\label{secII}}

We consider a mixture of a single-component bosonic superfluid and a fermionic superfluid
paring between two spin components, which rotates around the $z$ axis with
the same rotation frequency $\Omega$ for both superfluids.
We consider the experimental situation of a large number of particles in realistic geometries. To find out the equilibrium vortex states
in the rotating Bose-Fermi superfluid mixtures,
we use a relatively computationally simple model, i.e. the orbital-free DFT \cite{sadh2008,adh2010,yon2011,wen2018,kho2021}.
In terms of two complex-valued order parameters $\Psi_b$ for condensed bosons  \cite{pet2008}
and $\Psi_p$ for condensed fermionic pairs \cite{leg2006}, the energy functional associated with Bose-Fermi
mixtures in a rotating frame of reference is written, within the mean-field approximation, as
\begin{equation}\label{func}
E=\int \mathcal{E}[\Psi_b, \Psi_p]\;d{\bf r},
\end{equation}
where the energy density is
%
\begin{eqnarray*}\label{enden}
   \mathcal{E}[\Psi_b, \Psi_p]&=&\frac{\hbar^2}{2m_b}|\nabla\Psi_b|^2+V_b({\bf r})|\Psi_b|^2+\frac{1}{2}g_{bb}|\Psi_b|^4-\Psi^{\dagger}_b\Omega L_z\Psi_b\\
                  &+&\frac{\hbar^2}{4m_f}|\nabla\Psi_p|^2+2V_f({\bf r})|\Psi_p|^2+\frac{6}{5}\epsilon_f|\Psi_p|^2\sigma(\eta)-\Psi^{\dagger}_p\Omega L_z\Psi_p\\
                  &+&g_{bf}|\Psi_b|^2|\Psi_p|^2.
\end{eqnarray*}
%
The following calculations are performed in 3D formalism with
${\bf r}=\{x,y,z\}$. Here $m_b(m_f)$ is the mass of a bosonic (fermionic) atom, and $L_z=i\hbar(y\partial/\partial x-x\partial/\partial y)$ is the $z$
component of the angular momentum. The disk-shaped trapping potentials acting on bosons and fermions are given
by $V_{b,f}({\bf r})=m_{b,f}[\omega^2_{b\perp,f\perp}(x^2+y^2)+\omega^2_{bz,fz}z^2]/2$.
The total numbers of bosonic and fermionic atoms are determined,
by $N_b=\int d{\bf r}\;n_b=\int d{\bf r}\;|\Psi_b|^2$ and $N_f=\int d{\bf r}\;n_f=2\int d{\bf r}\;|\Psi_p|^2$, respectively.

Although it does not incorporate fermionic degrees of freedom, the orbital-free
approach in Eq.\;(\ref{func}) has computational advantage, and very recently has been used to be as a benchmark for experimental
observations \cite{raj2023,pac2022,fang2021,die2021}. Furthermore, by comparing with the fully microscopic theory \cite{bul2014,mmf2014,hos2022,wen2013},
the orbital-free DFT has been proved to be a good description for static properties and low-frequency linear dynamics, in
which pair-breaking effects play a negligible role \cite{mmf2014}.

The quantities $g_{bb}=4\pi\hbar^2a_{bb}/m_b$ and $g_{bf}=4\pi\hbar^2a_{bf}(m_b+m_f)/(m_bm_f)$ are the bosonic intraspecies and the boson-fermion interspecies interaction constants \cite{sadh2008}.
In contrast, the strength of the two-spin fermionic interaction in the BCS-BEC crossover is characterized by
the equation of state $\mu(n_f)= \partial [n_f\frac{3}{5}\epsilon_f\sigma(\eta)]/\partial n_f=\epsilon_f[\sigma(\eta)-(\eta/5)\partial \sigma(\eta)/\partial \eta]$, depending on
the Fermi energy $\epsilon_f=(\hbar k_f)^2/(2m_f)$ and the interaction parameter $\eta=1/(k_fa_f)$, with
the Fermi wave vector  $k_f=(3\pi^2n_f)^{1/3}$ and scattering length of fermions $a_f$.
$\sigma(\eta)$ is the fitting function that is a Pad$\acute{e}$-type parametrization of the experimental data \cite{nav2010} for a two-component Fermi gas at
zero temperature in the BCS-BEC crossover. To obtain a further analysis, we treat the equation of state by a polytropic approximation \cite{mani2005,wen2010}
\begin{subequations}\label{equst}
\begin{eqnarray}
\label{equf} \mu(n_f)&=&\mu^0(\frac{n_f}{n_0})^{\gamma} \\
\label{gamma} \gamma \equiv \gamma(\eta^0)=(\frac{n_f}{\mu}\frac{\partial \mu}{\partial n_f})|_{\eta=\eta^0}&=&
\frac{\frac{2}{3}\sigma (\eta^0)-\frac{2\eta^0}{5}\sigma^{\prime
}(\eta^0)+\frac{{\eta^0}^{2}}{15}\sigma^{\prime \prime}(\eta^0)}{\sigma (\eta^0)-\frac{\eta^0}{5}%
\sigma^{\prime }(\eta^0)},
\end{eqnarray}
\end{subequations}
where $\gamma$ is an effective polytropic index. The reference particle number density $n_0=(2m_f\epsilon^0_f)^{3/2}/(3\pi^2\hbar^3)$
is taken to be the density of the noninteracting Fermi gas at trap center, with the Fermi
energy $\epsilon^0_f=(\hbar k^0_f)^2/(2m_f)=\hbar(3N_f\omega^2_{f\perp}\omega_{fz})^{1/3}$, and
the reference chemical potential is $\mu^0=\epsilon^0_f[\sigma(\eta^0)-(\eta^0/5)\partial \sigma(\eta^0)/\partial \eta^0]$
with $\eta^0=1/(k^0_fa_f)$ and $k^0_f=(3\pi^2n_0)^{1/3}$ \cite{wen2010}.

Minimizing the energy functional with respect to variations of $\Psi_b$ and $\Psi_p$, and introducing the chemical potentials
$\mu_b$ and $\mu_p$ to fix the particle numbers $N_b$ and $N_f$, yields the following equations
\begin{subequations}\label{cporE}
\begin{eqnarray}
\label{ordB} \mu_b\Psi_b &=& \left[-\frac{\hbar^2 \nabla^2}
{2m_b}+V_{b}+ g_{bb}|\Psi_b|^2+g_{bf}|\Psi_p|^2-\Omega L_z\right]\Psi_b \\
\label{ordF} \mu_p \Psi_p &=& \left[-\frac{\hbar^2 \nabla^2}
{4m_f}+2V_f+ 2\mu(n_{f})+g_{bf}|\Psi_b|^2-\Omega L_z\right]\Psi_p,
\end{eqnarray}
\end{subequations}
with $\nabla^2=\partial^2/\partial x^2+\partial^2/\partial y^2+\partial^2/\partial z^2$.
We introduce the energy, time and length scales,
given by $\hbar\omega_{b\perp}$, $\omega^{-1}_{b\perp}$ and $\ell_{b\perp}=\sqrt{\emph{}\hbar/(m_b\omega_{b\perp})}$, respectively.
The order parameters are normalized by the atomic numbers in 3D as $\Psi_b\rightarrow \sqrt{N_b}\Psi_b/\ell^{3/2}_{b\perp}$
and $\Psi_p\rightarrow \sqrt{N_f/2}\Psi_p/\ell^{3/2}_{b\perp}$, and $\int d{\bf r} |\Psi_{b,p}|^2=1$.
To find the stationary states of Eqs.\;(\ref{cporE}), we use the imaginary time propagation
of the time-dependent version of Eqs.\;(\ref{cporE}) after sufficient convergence \cite{kas2003,kas2009}.
The time-dependent version of Eqs.(\ref{cporE}) takes the following dimensionless form
\begin{subequations}\label{icporE}
\begin{eqnarray}
\label{iordB} i\frac{\partial  \Psi_b}{\partial t} &=& \left[-\frac{1}
{2}\nabla^2+\tilde{V}_b+ u_{b}|\Psi_b|^2+u_{fb}|\Psi_p|^2-\tilde{\Omega} \tilde{L}_z\right]\Psi_b, \\
\label{iordF} i\frac{\partial  \Psi_p}{\partial t} &=& \left[-\frac{\alpha}
{2} \nabla^2+\tilde{V}_f+u_f|\Psi_p|^{2\gamma}+u_{bf}|\Psi_b|^2-\tilde{\Omega} \tilde{L}_z\right]\Psi_p,
\end{eqnarray}
\end{subequations}
where the mass ratio is defined by $\alpha=m_b/(2m_f)$. Here the rotation frequency
is $\tilde{\Omega}=\Omega/\omega_{b\perp}$ and $\tilde{L}_z=L_z/\hbar$, the trapping potentials
for bosons and fermions are $\tilde{V}_b=(x^2+y^2+\omega^2_{bz}z^2/\omega^2_{b\perp})/2$ and
$\tilde{V}_f=[\omega^2_{f\perp}(x^2+y^2)+\omega^2_{fz}z^2]/(2\alpha\omega^2_{b\perp})$,
and the dimensionless parameters for the intra- and interspecies interactions are $u_b=4\pi N_ba_b/\ell_{b\perp}$,
$u_f=2\mu^0(N_f/2n_0\ell^3_{b\perp})^{\gamma}/(\hbar\omega_{b\perp})$, $u_{fb}=2\pi N_fm_ba_{bf}/(m_{bf}\ell_{b\perp})$,
and $u_{bf}=4\pi N_bm_ba_{bf}/(m_{bf}\ell_{b\perp})$, respectively.

We consider the parameters of the experiment \cite{yao2016},
in which vortex lattices are created in a rotating $^{41}$K-$^6$Li superfluid mixture with an imbalanced mass $\alpha=3.4$.
The bosons and fermions feel different radial frequencies $\omega_{b\perp}=2\pi\times 20$$\rm{Hz}$ ($\omega_{f\perp}=2\pi\times 40$$\rm{Hz}$)
and axial frequencies $\omega_{bz}=2\pi\times 85$${\rm Hz}$ ($\omega_{fz}=2\pi\times 237$${\rm Hz}$) of the disk-shaped trapping potentials.
In order to enhance the contrast of vortices, the particle numbers are chosen $N_b=1\times10^4$ and $N_f=2\times10^5$, respectively, both to be
an order of magnitude smaller than the experiment. The scattering lengths for bosons is $a_b=60.5a_0$
with $a_0$ being the Bohr radius, and the boson-fermion scattering length $a_{bf}$ varies positive over a large range to realize a
miscibility-immiscibility transition.

Within this framework, we consider the problem by solving two coupled nonlinear Schr\"{o}dinger equations (\ref{icporE}),  with the polytropic equation of state Eq.(\ref{equf})
treating exactly several important regimes of interacting Fermi superfluids.
Therefore, one can study the properties of a Bose-Fermi mixture through the BCS-BEC crossover in a unified way,
i.e. from a mixture of a weakly interacting Bose superfluid and a strongly interacting Fermi superfluid to
a weakly interacting two-component BECs. For instance, at the unitary point ($\eta^0=0$),
the equation of state is characterized by the parameters $\mu^0=0.412 \epsilon^0_f$ and $\gamma=2/3$, which
takes a universal density dependence $\mu^0/n^{\gamma}_0=0.412\hbar^2(3\pi^2)^{2/3}/(2m_f)$ \cite{zwe2012}. In the deep BEC regime ($\eta^0=6, a_f=1168a_0$)
characterized by $\mu^0=0.01 \epsilon^0_f$ and $\gamma=1.01$, the equation of state in terms of the above parameters
takes $2\mu^0/n^{\gamma}_0=0.66g_{M}$, slightly different from the mean-filed interaction $g_M=4\pi\hbar^2a_M/(2m_f)$ of
a BEC with the molecule-molecule scattering length $a_M=0.6a_f$.
It is because that the used equation of state includes the beyond-mean-field correction \cite{nav2010}.
In the BEC limit ($\eta^0=16(70), a_M=263(61)a_0$), the equation of state characterized by $\mu^0=0.004(0.0009) \epsilon^0_f$ and $\gamma=1.003(1.0003)$
can reproduce the expected mean-field interaction $2\mu^0/n^{\gamma}_0=0.9(1.0)g_{M}$ well.

For initial conditions, we consider the ones with a single vortex aligned with $z$ axis
at the center, modulated by a random phase at different space points \cite{ram2019}
\begin{subequations}\label{iniw}
\begin{eqnarray}
\label{iniwb} \Psi_{b,0}=(\frac{\omega_{bz}}{\pi\omega_{b\perp}})^{\frac{1}{4}}\frac{x+iy}
{\sqrt{\pi}}\exp[-\frac{x^2+y^2}{2}-\frac{\omega_{bz}z^2}{2\omega_{b\perp}}+2\pi i \Re(x,y)] \\
\label{iniwf} \Psi_{p,0}=(\frac{\omega^4_{f\perp}\omega_{fz}}{\alpha^5\pi\omega^5_{b\perp}})^{\frac{1}{4}}\frac{x+iy}
{\sqrt{\pi}}\exp[-\frac{\omega_{f\perp}(x^2+y^2)}{2\alpha\omega_{b\perp}}-\frac{\omega_{fz}z^2}{2\alpha\omega_{b\perp}}+2\pi i \Re(x,y)].
\end{eqnarray}
\end{subequations}
$\Re(x,y)$ is a randomly generated number distributed uniformly between 0 and 1.
The included random phase term breaks the underlying symmetries and prevents the simulation from getting stuck in any metastable states.
A combination of angular harmonics with the randomly generated numbers
has been successfully used to generate vortex lattices in dipolar two-component BECs \cite{kum2017}.
The numerical method is based on the split-step Crank-Nicolson scheme \cite{mur2009}. Vortices
arrange themselves inside the trap, and the system is closer to the equilibrium configuration for long times.
After each time step of computations, the wavefunctions for both superfluids are renormalized to one.
Imaginary-time propagation is conducted until the desired precision is reached for the energy or chemical potential \cite{ram2019}.
The numerical simulations are conducted in a grid with a maximum of $600\times600\times48$ points along the $x$, $y$
and $z$ directions respectively, with a spatial step of $0.1$ in both $x$ and $y$ directions, and $0.5$ in the $z$ direction,
and a time step of $0.001$. To speed up the calculation, the programs are parallelized
using Open Multi-processing (OpenMP) interface \cite{sat2015} run on supercomputing system.

\section{vortex lattices in strongly interacting Fermi superfluids}\label{secIII}

\begin{figure}[h]
\begin{center}
\includegraphics[width=6in]{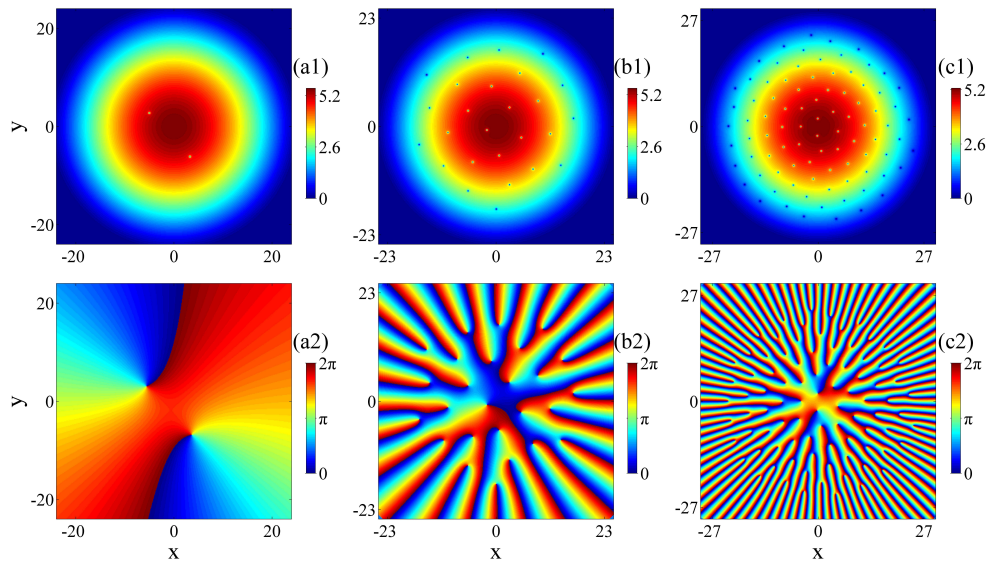}
\end{center}
\caption{\label{fig1} The formations of vortex lattices in a strongly interacting Fermi superfluid
as the rotation frequency from (a) $\Omega=0.05\omega_{f\perp}$, (b) $0.15\omega_{f\perp}$, to
(c) $0.7\omega_{f\perp}$. Shown in panels (a1)-(c1) are the corresponding cross-sectional densities (in units of $10^{-4}$) at $z=0$ plane, and (a2)-
(c2) are the corresponding cross-sectional phases.}
\end{figure}

Vortex lattices in a strongly interacting Fermi superfluid have been studied extensively
by using the orbital-based DFT \cite{dav2004,sim2015,kop2021}.
In particular, by solving non-uniform BdG equations in a local phase density approximation,
vortex lattices  in the BCS-BEC crossover in a real 3D trap
have been investigated \cite{sim2015}, addressing a comparison with experimental data.
In this section, we first study the formation of vortex lattices in Fermi superfluids alone in the BCS-BEC crossover.
Because of a similar configuration, it is of interest and necessary to compare our results with the full microscopic theory \cite{sim2015}.
In Fig.~\ref{fig1} we present our results for the vortex lattices
in the unitary Fermi superfluid ($\eta^0=0$) as a function of the rotation frequency $\Omega$.
It should be noticed that the numerical
simulations are actually carried out in enough larger cuboid computational domains to avoid the boundary
effects, and the extra boundary areas are cut for a better presentation.
One can find the generated vortices emerging in the cross-sectional
density profiles $n_f(x,y,z=0)$, verified them
in corresponding cross-sectional phases, i.e. ${\rm arctan} [{\rm Im}\Psi_p(x,y,z=0)/{\rm Re}\Psi_p(x,y,z=0)]$,
by the varying value from $0$ to $2\pi$.

\begin{figure}[h]
\centering
\includegraphics[width=4.5in]{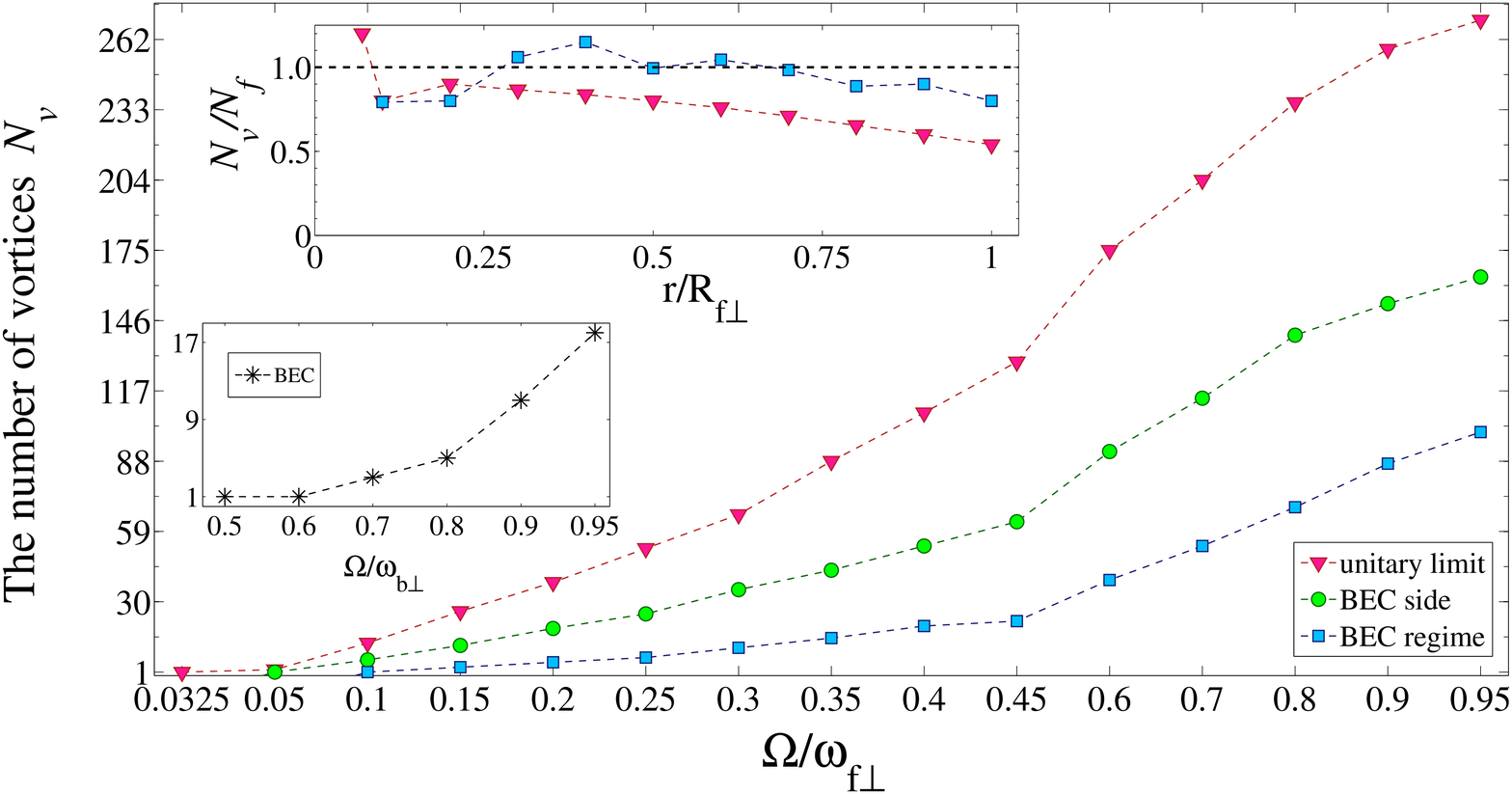}
\caption{\label{fig2} The vortex numbers as a function of the rotation frequency $\Omega$
for the different interaction regimes: the unitary limit ($\eta^0=0$), the BEC side ($\eta^0=1$),
and the BEC regime ($\eta^0=6$). Lower inset: the behavior of the vortex number in the weakly
interacting Bose superfluid. Upper inset: the ratio between the number of vortices $N_v$ obtained numerically
and the prediction $N_f$ from Feynman's theorem
as a function of the radius $r=\sqrt{x^2+y^2}$ for $\Omega/\omega_{f\perp}=0.95$.}
\end{figure}

In regard to the properties of vortex lattices in a rotating strongly interacting Fermi superfluid, one is the dependence of the
vortex number $N_v$ on the rotation frequency $\Omega$. Fig.~\ref{fig2} shows $N_v$ versus $\Omega$ in the different interaction regimes,
and the case of the weakly interacting Bose superfluid is also plotted in the lower inset of Fig.~\ref{fig2} for comparison. It is seen that
the number of vortices follows a linear dependence on $\Omega$ at the low rotation frequency, and the rate of increase is larger and
larger as $\Omega$ increases. At the large rotation frequency, when entering into the strongly interacting regimes (i.e. $\eta^0=0$ and 1),
however, the increase of the vortex number is suppressed displaying a nonlinear dependence.
At the very high frequencies, a rotating superfluid mimics rigid
body rotation  with the average curl of the velocity field $\nabla\times \vec{v}=2\vec{\Omega}$.
The areal density of vortices $n_v$ obeys the Feynman's relation \cite{fey1955} applied to a uniform vortex distribution
\begin{equation}\label{meandensity}
n_v=\frac{2m_f\Omega}{\pi\hbar},
\end{equation}
which has a factor 2 for a Fermi superfluid compared with that of a Bose superfluid with the same value of
atomic mass. In the upper inset of Fig.~\ref{fig2} for $\Omega/\omega_{f\perp}=0.95$, we shows the ratio between the number of vortices $N_v(r)$
obtained numerically within a circle with the radius $r=\sqrt{x^2+y^2}$, and
the corresponding results $N_f(r)=n_v\pi r^2$ expected from Feynman's theorem. One can find
that they are in good agreement near the trap center, but
the density inhomogeneity from the numerical results reduces the vortex density away from the center \cite{sim2015}. Up to the
boundary of the superfluid ($r/R_{f\perp}=1$), the ratio $\frac{N_v}{N_f}=\frac{270}{502}=0.54$ in the
unitary is smaller than $\frac{N_v}{N_f}=\frac{100}{125}=0.8$ in the BEC regime. It suggests that the inhomogeneity correction
in the strongly interaction regime is larger, and explains the increase suppression and nonlinear trend of the vortex numbers
at the large rotation frequency.

In Ref.\;\cite{sim2015} based on the BdG calculations, the vortex number is obtained $N_v=\{85,112,137\}$ for  $\Omega=\{0.4,0.6,0.8\}\omega_{f\perp}$ in the unitary limit,
which is smaller than our results $N_v=\{108,175,236\}$ as shown in Fig.~\ref{fig2}. In addition, the maximum of $N_v$ for a fixed $\Omega$ is
at the BEC side of the crossover, shifting towards the BEC side as $\Omega$ increases. This is in contrast to our results showing a monotonic
increase of $N_v$ from the BEC regime to the unitary limit.
Such discrepancies are attributed to that the microscopic theory can account for the filling of fermionic vortex cores with a normal
component \cite{kop2021,aur2003,hui2006}, and the density depletion at the vortex core is not completed around the unitary limit even at zero-temperature. Our formulation only in terms of the order parameter misses the normal state. As a result, a vanishing order parameter yields a vanishing density \cite{wen2013,zhang2020},
which magnifies the contrast of the vortex core and the vortex number. In the BEC regime ($\eta^0=1$) where the density depletion of the vortex core
is also completed from the microscopic theory, therefore our results $N_v=\{53,92\}$ for $\Omega=\{0.4,0.6\}\omega_{f\perp}$ are
in better agreement with the microscopic theory $N_v=\{60,100\}$ \cite{sim2015}.

The other property is the critical frequency $\Omega_{c}$ of the vortex nucleation in the strongly interacting Fermi superfluids \cite{bru2001,hui2006}.
The thermodynamic critical frequency can be calculated as $\Omega_{c}=(E_1-E_0)/\langle L_z \rangle$ analytically,
where $E_1$ and $E_0$ are the energy of the single-vortex state and the vortex-free energy, respectively,
and $\langle L_z\rangle $ is the mean angular momentum of a vortex state.
We have previously obtained the extra energy for per unit length of a uniform Fermi system \cite{zhang2020}, with a single quantum
of circulation lying long the axis of a cylinder of radius $R_{f\perp}$.
Dividing the extra energy by the angular momentum yields the characteristic frequency of the first
vortex nucleation
\begin{equation}\label{crv}
\Omega_{c}=\frac{\hbar}{2m_f R^2_{f\perp}} \ln (1.464 \gamma^{\frac{2}{5}} \frac{R_{f\perp}}{\xi_f}),
\end{equation}
with the Thomas-Fermi (TF) radius $R_{f\perp}=\sqrt{2\mu_f/(2m_f\omega^2_{f\perp})}$ and the coherence length
$\xi_f=\hbar/\sqrt{4m_f\mu_f}$ evaluated by the central density. They are both determined by the chemical potential
$\mu_f=\hbar\omega_{b\perp}[\omega_{fz}(\omega^2_{f\perp}/(2\alpha\omega_{b\perp}))^{\frac{3}{2}}u_f^{\frac{1}{\gamma}}
\Gamma(\frac{1}{\gamma}+\frac{5}{2})/(\omega_{f\perp}\Gamma(\frac{1}{\gamma}+1)\pi^{\frac{3}{2}})]^{2\gamma/(2+3\gamma)}$ of the Fermi superfluid.
In the BEC limit ($\gamma=1$), Eq.\;(\ref{crv}) can reproduce the result $\Omega_{c}=\hbar/(m_bR^2_{b\perp}) \ln(1.464 R_{b\perp}/\xi_b)$
for the Bose superfluid, with $R_{b\perp}=\sqrt{2\mu_b/(m_b\omega^2_{b\perp})}$,
$\xi_b=\hbar/\sqrt{2m_b\mu_b}$, and
$\mu_b=\hbar \omega_b[15u_b\omega_{bz}/(16\sqrt{2}\pi \omega_{b\perp})]^{5/2}$.

For our chosen parameters, the critical frequencies from Eq.(\ref{crv}) for
a uniform rotating system are given by $\Omega_{c}=\{0.016, 0.03, 0.06\}\omega_{f\perp}$ for
the unitary limit ($\eta^0=0$), the BEC side ($\eta^0=1$) and
the BEC regime ($\eta^0=6$), respectively, which are smaller than the numerical results $\Omega_{c}=\{0.0325,0.05,0.1\}\omega_{f\perp}$ (see Fig.~\ref{fig2}).
This is because that the nonuniform density in the axisymmetric trap reduces the total angular momentum
relative to that of a uniform system. In the case of BEC, by taking into account inhomogeneity \cite{pet2008}
the critical frequency is $\Omega_{c}=5\hbar/(2m_bR^2_{b\perp})\ln(0.671R_{b\perp}/\xi_b)=0.42\omega_{b\perp}$,
larger than $\Omega_c=0.2\omega_{b\perp}$ from Eq.\;(\ref{crv}),
and in good agreement with the numerical result $\Omega_c=0.5\omega_{b\perp}$ (see the lower inset of Fig.~\ref{fig2}).
The monotonic increase of the critical frequency from the unitary limit to the BEC side of the crossover
is also found from the BdG calculations. The critical frequency $\Omega_{c}=0.069\omega_{f\perp}$
in the unitary limit is reported \cite{sim2015}, which is a little larger than ours traced to the same reason as before.

\section{The impact of boson-fermion interaction}{\label{secIV}}


\begin{figure}[h]
\centering
\includegraphics[width=5.5in]{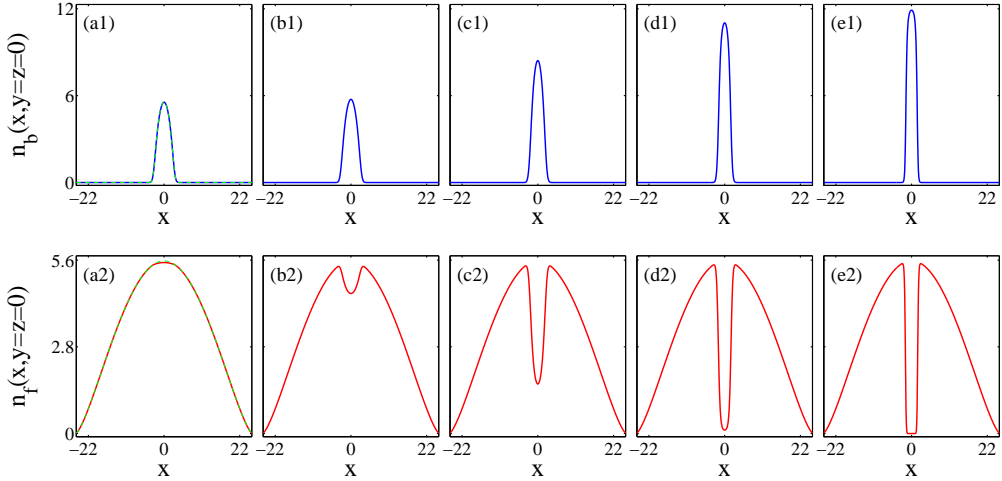}
\caption{\label{fig3} The nonrotating densities of the Bose (upper panels, in units of $10^{-2}$) and
Fermi (lower panels, in units of $10^{-4}$) superfluids as a function of $x$. The boson-fermion scattering length
increases from (a) $\tilde{a}_{bf}=a_{bf}/(60.9a_0)=0.05$, (b) $1$, (c) $3$, (d) $4$ to
(e) $6$. The densities of the Bose and Fermi superfluids without the interaction ($\tilde{a}_{bf}=0$)
are also drawn as dashed lines in panels (a1) and (a2), respectively, for comparison.}
\end{figure}

\begin{figure}[h]
\centering
\includegraphics[width=5in]{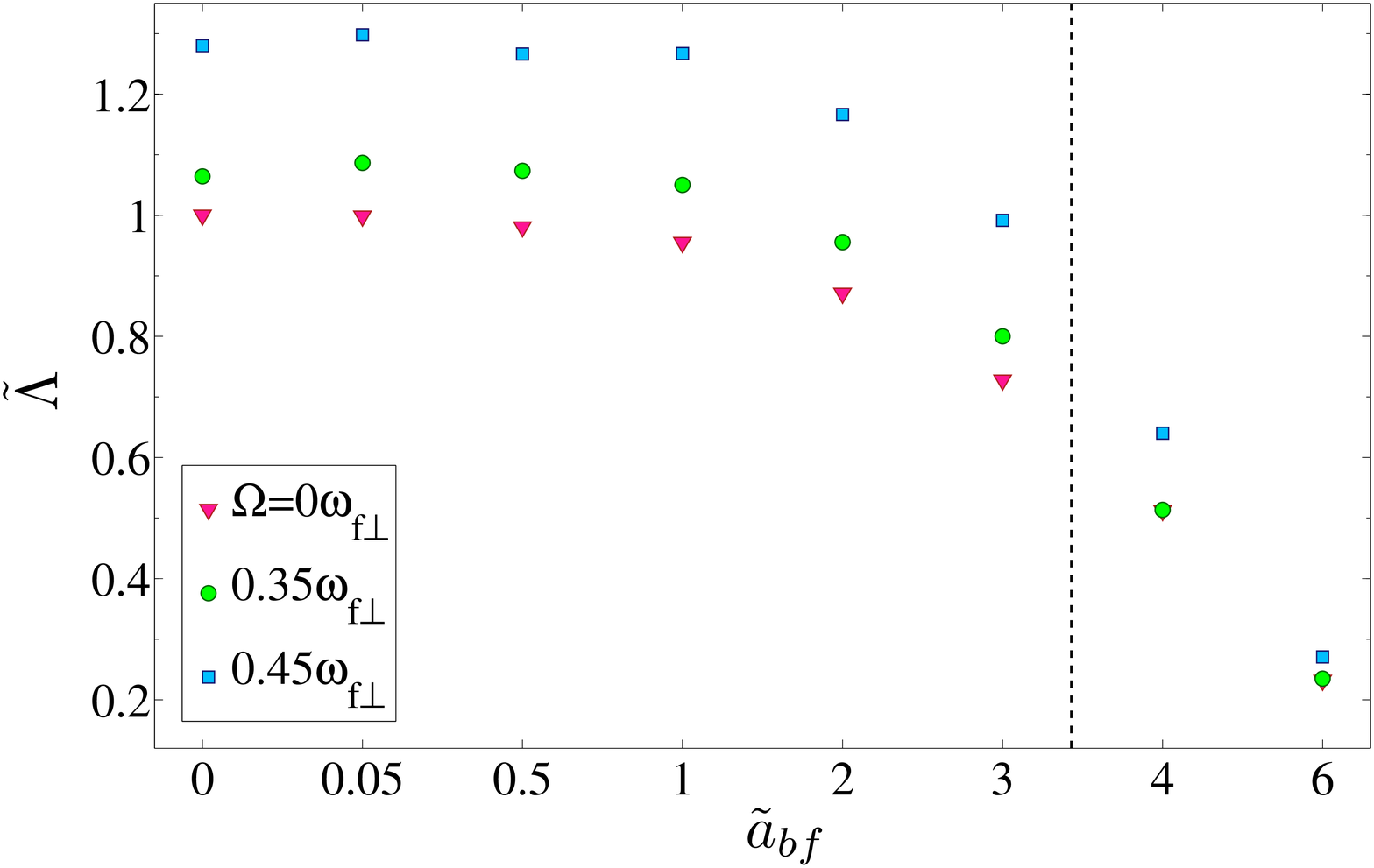}
\caption{\label{fig4} The scaled overlap parameter $\tilde{\Lambda}=\Lambda/\Lambda_0$
of the Bose-Fermi superfluid mixture as a function of $\tilde{a}_{bf}$ for various rotation frequencies.
The value $\Lambda_0=0.135$ is for the case without interspecies interaction ($\tilde{a}_{bf}=0$)
and rotation ($\Omega=0$). The analytical prediction $\tilde{a}_{bf}=3.4$ of
the miscible-immiscible transition without rotation
is marked with the vertical dashed line.}
\end{figure}

In this section, we next investigate the effects of repulsive boson-fermion interactions on the rotating Bose-Fermi superfluid mixtures, focusing on the unitary limit ($\eta^0=0$) in
the experimentally accessible parameters. To evaluate the strength of the repulsive boson-fermion interaction, in Fig.~\ref{fig3} we
present the density distributions of nonrotating Bose and Fermi superfluids ($\Omega=0$) for various boson-fermion interactions. As a reference,
we introduce a dimensionless scattering length $\tilde{a}_{bf}=a_{bf}/(60.9a_0)$ scaled by the case studied in the experiment \cite{yao2016}.
A relevant characteristic of coupled mixtures is the miscibility of the components \cite{bho2008}.
In the miscible phase, the densities of two components overlap with each other;
whereas, they get spatially separated in immiscible phase. For a weak repulsive interspecies interaction ($\tilde{a}_{bf}=0.05$)
in Fig.~\ref{fig3}(a), it is shown that the density distributions of the Bose and Fermi superfluids are almost identical to
the uncoupled densities ($\tilde{a}_{bf}=0$) by the dashed lines.
Increasing the repulsive interaction $(\tilde{a}_{bf}=1)$ in Fig.~\ref{fig3}(b) which is for the experimental case,
the density of the Fermi superfluid reduces at the center pronouncedly, but the mixture is still overlapping in the miscible phase.
Further increasing the interaction $(\tilde{a}_{bf}=6)$ in Fig.~\ref{fig3}(e) to induce immiscibility
transition, there is a shell structured geometry, in which
the bosonic atoms occupy the small central region as the core-part, and the density of the fermions is zero
at the center forming the large shell-part.

A parameter to measure the spatial overlapping between densities of the components
can be given by
\begin{equation}\label{of}
\Lambda=\int d{\bf r} \sqrt{|\Psi_b|^2|\Psi_p|^2},
\end{equation}
where the order parameters $\Psi_b$ and $\Psi_p$ are both normalized to one.
Even without interspecies interaction, the overlap parameter in our case is merely up to $\Lambda_0=0.135$.
In the presence of the interaction $\tilde{a}_{bf}=1$, we then obtain a smaller value of $\Lambda=0.129$
indicating partial overlapping for the ratio $\tilde{\Lambda}=\Lambda/\Lambda_0=0.95$. Fig.~\ref{fig4} shows the
scaled overlap parameter $\tilde{\Lambda}$ as a function of $\tilde{a}_{bf}$.
It is clearly that the system undergoes a miscible-immiscible transition in the parameter range of $\tilde{a}_{bf}=3\sim4$ corresponding to $\Lambda=0.1\sim 0.07$.

The energy density of a homogenously mixed phase of Bose-Fermi superfluid mixture can be obtained
from Eq.\;(\ref{func}) by neglecting the kinetic energy terms,
that is  $\mathcal{E}_r=3\hbar^2(3\pi^2)^{\frac{2}{3}}n^{\frac{5}{3}}_f\sigma(\eta)/(10m_f)+
g_{bb}n^2_b/2+g_{bf}n_bn_f/2$. A condition for miscibility is that the Hessian matrix
of $\mathcal{E}_r$ is positive semidefinite \cite{fla2007}, i.e. $(\partial^2\mathcal{E}_r/\partial n^2_b)
(\partial^2\mathcal{E}_r/\partial n^2_f)-(\partial^2 \mathcal{E}_r/\partial n_b \partial n_f)^2>0$.
The solution of this inequality gives the parameter regime of boson-fermion scattering length
\begin{equation}\label{msp}
a^2_{bf}<\frac{3m^2_{bf}a_{bb}(3\pi^2)^{\frac{2}{3}}}{10\pi m_b m_f}\frac{\partial^2}{\partial n^2_f}[n^{\frac{5}{3}}_f\sigma(\eta)],
\end{equation}
where the homogeneous mixed phase is energetically stable. In the unitary limit $\sigma(\eta)$ is taken as the universal factor $\xi=0.41$  \cite{nav2010}, and $n_f$ is approximated by the Fermi density at the trap center without the interspecies interaction. The critical value $\tilde{a}^c_{bf}=3.4$ for miscibility of the Bose and Fermi superfluids is predicted analytically, which is denoted by the vertical dashed line. Fig.~\ref{fig4} also compares the behavior of the miscibility as a function of the boson-fermion interaction under different rotations. One can find that the miscibility of the mixtures enhances due to the centrifugal force
and increases as the rotation frequency. In addition, as the interspecies interaction increases, the discrepancies of the miscibility
for different rotation frequencies in the phase-separated regime are narrowed.

Here we start from a very small value of the rotation frequency, and examine the effect of boson-fermion interactions on
the vortex nucleation. We find that the boson-fermion interaction decreases the critical frequency of the Fermi superfluid
from $\Omega_c=0.0325\omega_{f\perp}$ to $0.031\omega_{f\perp}$  slightly. In the presence of
a very weak boson-fermion interaction $\tilde{a}_{bf}=0.05$, the first vortex appears at $\Omega=0.031\omega_{f\perp}$.

\begin{figure}[h]
\centering
\includegraphics[width=7in]{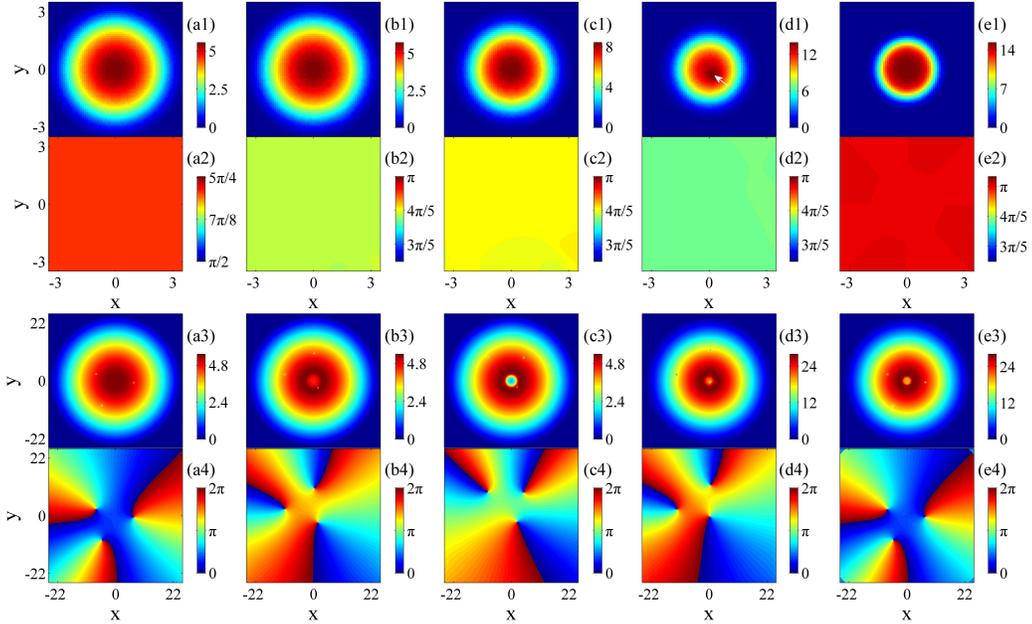}
\caption{\label{fig5} Vortex-lattice structures
in the slowly rotating Bose-Fermi superfluid mixtures ($\Omega=0.05\omega_{f\perp}$),
for (a) $\tilde{a}_{bf}=0.05$, (b) $1$, and (c) $3$ in the miscible regime.
Panels (a1)-(c1) show the cross-sectional densities (in
units of $10^{-2}$) of the Bose superfluid at $z=0$ plane, and (a3)-(c3) show the cross-sectional
densities (in units of $10^{-4}$) of the Fermi superfluid. In contrast,
(d) $\tilde{a}_{bf}=4$ and (e) $6$ correspond to the immiscible phases.
(d1) and (e1) show the integrated densities $\int n_b(x,y,z)dz$ (in units of $10^{-2}$) of the Bose superfluid,
(d3) and (e3) are the integrated densities $\int n_f(x,y,z)dz$ (in units
of $10^{-4}$) of the Fermi superfluid. Panels (a2)-(e2) and (a4)-(e4) correspond to the cross-sectional phases of the Bose
and Fermi superfluids at $z=0$ plane, respectively.}
\end{figure}

Fig.~\ref{fig5} illustrates
the case of slowly rotating $\Omega=0.05\omega_{f\perp}$.
Fig.~\ref{fig5}(a) corresponds to a very weak boson-fermion strength $\tilde{a}_{bf}=0.05$.
By comparing with Fig.~\ref{fig1}(a) only two vortices in the single Fermi superfluid,
one can see that one extra vortex emerges instantaneously
in Fig.~\ref{fig5}(a3) even in the presence of the weak interaction.  With further increasing $\tilde{a}_{bf}=1$ in Fig.~\ref{fig5}(b) and
$\tilde{a}_{bf}=3$ in Fig.~\ref{fig5}(c) that are both still in the miscible regime, the vortex number keeps invariant but the vortex configurations evolves.
As the central density becomes more depressed, the three vortices are attracted spirally to the center areas, which can be seen more clearly from the corresponding phases in Fig.~\ref{fig5}(a4)-(c4).
In this experimental relevant system, the trap frequencies for bosons and fermions are not equal, i.e. $\omega_{f\perp}/\omega_{b\perp}=2$. The rotation frequency is
given by $\Omega=0.05\omega_{f\perp}=0.1\omega_{b\perp}$, which is much smaller than the critical frequency $0.5\omega_{b\perp}$ of the Bose superfluid. The Bose superfluid is thus without vortex nucleation, and the density profiles shown
in Fig.~\ref{fig5}(a1)-(c1) are like nonrotation and the corresponding phases shown in Fig.~\ref{fig5}(a2)-(c2).
The numerical results are consistent with the experimental observation that the boson-fermion interaction leads to
unexpected vortex formation compared to a single Fermi superfluid and increases the vortex number.

Instead of the cross-sectional densities shown in Fig.~\ref{fig5}(a)-(c), Fig.~\ref{fig5}(d) and \ref{fig5}(e)
present the integrated densities of the cases in the immiscible regime.
It is because that in the immiscible state the cross-sectional density of the Fermi superfluid
at the center is zero and the vortex is invisible. For $\tilde{a}_{bf}=4$, we observe the formation of the coreless vortex in Fig.~\ref{fig5}(d).
One of vortices in the Fermi superfluid enters into the overlapping center area (Fig.~\ref{fig5}(d3)), which creates a density peak
in the Bose superfluid seen as a dark spot in Fig.~\ref{fig5}(d1) (denoted by the arrow).
The coreless vortex was first experimentally created in two-component BECs \cite{mat1999}.
In terms of a pseudospin representation \cite{kas2005}, an axisymmetric vortex is interpreted as skyrmions, in which the vortex core of one component
is filled with the other nonrotating component, while a nonaxisymmetric one is regarded as meron pairs, in which each component has one off-centered vortex \cite{keni2005}.
More recently, the static and dynamical properties of  massive vortices (coreless vortex) have been studied by means of a massive point-vortex model \cite{Ric2020,Ric2021}.
For a larger value $\tilde{a}_{bf}=6$ in Fig.~\ref{fig5}(e) where the Bose and Fermi superfluids are well separated,
however, the coreless vortex disappears since the vortex is repelled from the overlapping center area of the Fermi superfluid (Fig.~\ref{fig5}(e3)).
Our results indicate that a coreless vortex state with distinct quantum statistics can be observed in the slowly rotating
Bose-Fermi superfluid mixtures around the miscible-immiscible transition.

Increasing the rotation frequency above the threshold for the appearance of the first vortex
in the Bose superfluid, we can study the interplay between the vortex lattices emerging  in two different superfluids.
We find that for a very weak interspecies interaction ($\tilde{a}_{bf}=0.05$) the critical frequency of the Bose superfluid
decreases $\Omega_c=0.45\omega_{b\perp}$ compared with $0.5\omega_{b\perp}$ for a single one.

\begin{figure}[h]
\centering
\includegraphics[width=7in]{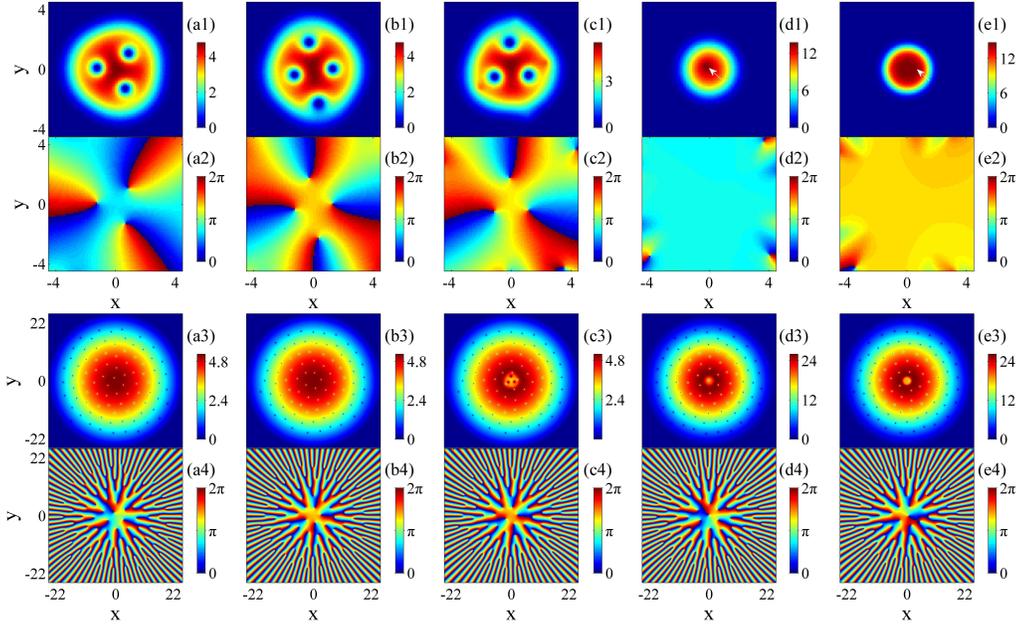}
\caption{\label{fig6} Vortex-lattice structures in the moderately rotating Bose-Fermi
superfluid mixtures ($\Omega=0.35\omega_{f\perp}$), for (a) $\tilde{a}_{bf}=0$, (b) $0.05$,
and (c) $2$ in the miscible regime, and (d) $\tilde{a}_{bf}=4$ and (e) $6$ in the immiscible regime.
As the same as in Fig.~\ref{fig5}, panels (a1)-(c1) and (d1)-(e1) show the cross-sectional and integrated densities
of the Bose superfluid, respectively, and correspondingly (a3)-(c3) and (d3)-(e3) present the Fermi superfluid.
Panels (a2)-(e2) and (a4)-(e4) are the corresponding cross-sectional phases.}
\end{figure}

In Fig.~\ref{fig6}, we present the structural variations of vortex lattices in
a moderately rotating Bose-Fermi superfluid mixture ($\Omega=0.35\omega_{f\perp}$) through the miscible-immiscible transition.
In the absence of the interaction $\tilde{a}_{bf}=0$ in Fig.~\ref{fig6}(a), the Bose and Fermi superfluids behave independently and the
vortex lattices in two superfluids are uncoupled. With the onset of the very weak interaction $\tilde{a}_{bf}=0.05$
in Fig.~\ref{fig6}(b), one extra vortex is found to appear immediately in the Bose superfluid, and the vortex number increases from 3 to 4
distinctly in Fig.~\ref{fig6}(b1). The interspecies interaction also affects the arrangement of the vortices in the Fermi superfluid, which distribute spirally from the center more regularly (Fig.~\ref{fig6}(b3)).
Continually turning $\tilde{a}_{bf}=2$ up in Fig.~\ref{fig6}(c1) still in the miscible regime,
we observe the annihilation of the vortex lattice with one of the vortices disappearing, and the vortex number is the same as that in Fig.~\ref{fig6}(a1) for no interaction.

\begin{figure}[h]
\centering
\includegraphics[width=7in]{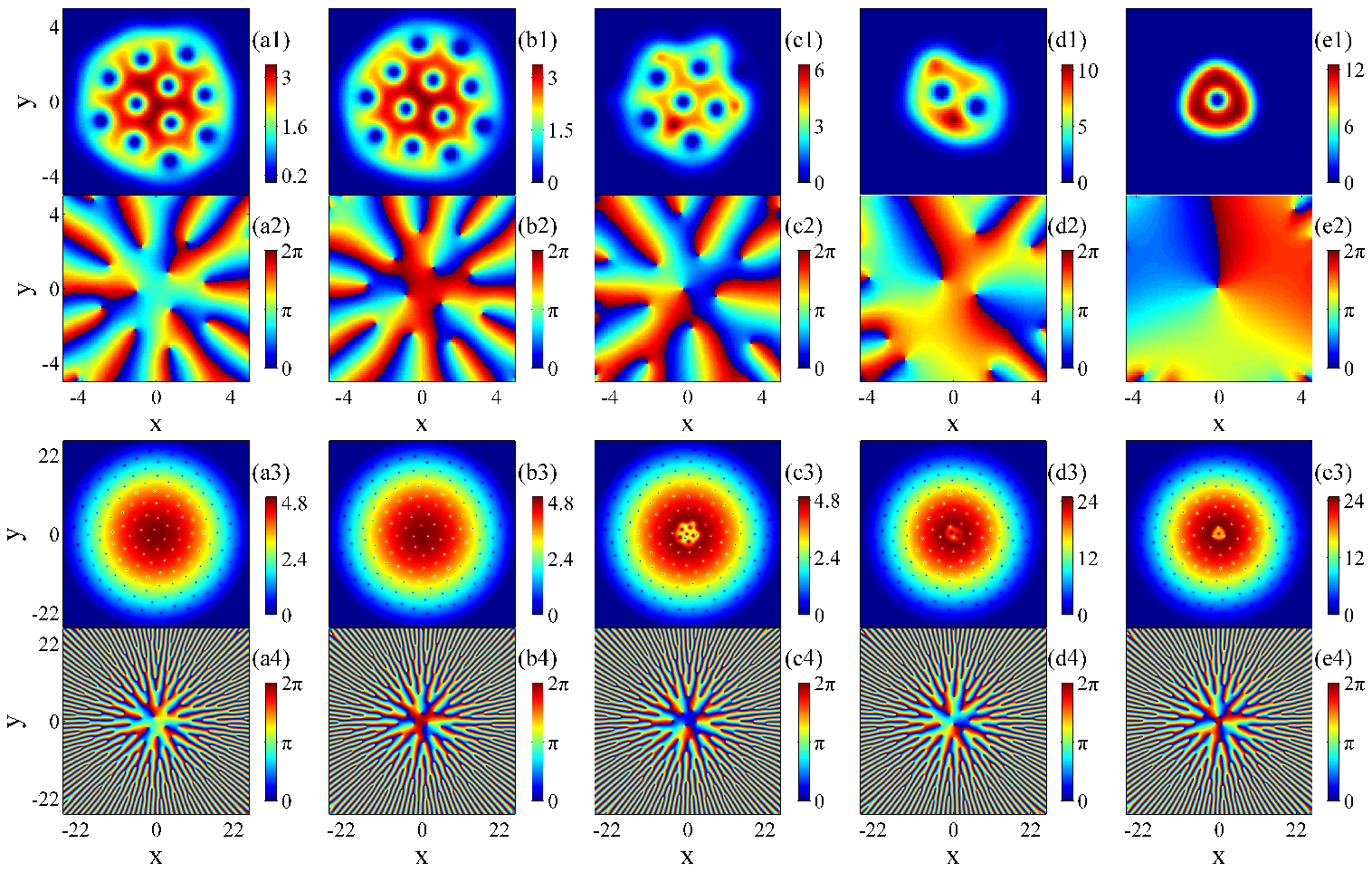}
\caption{\label{fig7}  The same case as Fig.~\ref{fig6}, but for a larger rotation $\Omega=0.45\omega_{f\perp}$,
and (a) $\tilde{a}_{bf}=0$, (b) $0.05$, (c) $3.0$, (d) $4.0$, and (e) $6.0$.}
\end{figure}

In the experiment a few unconventional behaviors are observed  \cite{yao2016}.  The number and lifetime of vortices
in the Fermi superfluid are greater than that in a single superfluid. In contrast, the effects of the boson-fermion interaction are less pronounced on
the Bose superfluid. However, our numerical results in Fig.~\ref{fig6}(a)-(c) indicate a different behavior of the vortex number of the Bose superfluid.
By comparing with Fig.\;4 of Ref.\;\cite{yao2016},
more vortices in the Bose superfluid are observed, and
the effects are measured at $\tilde{a}_{bf}=1$ smaller than $2$ where we find the decrease of the vortex number.
The particle number of the bosons in the experiment is an order of magnitude larger than ours.
Therefore, more particles result in more vortices and enhance the miscible-immiscible transition
that is $\tilde{a}^c_{bf}=2.8$ from Eq.\;(\ref{msp}).

For the case of $\tilde{a}_{bf}=4$ in Fig.~\ref{fig6}(d), the Bose-Fermi superfluid mixture in the immiscible state and no vortices are left to be
visible in the Bose superfluid (see Fig.~\ref{fig6}(d1) and the corresponding phase Fig.~\ref{fig6}(d2)). Similar to Fig.~\ref{fig5}(d),
the vortex-lattice structure is also featured by a coreless vortex.
But differently from its disappearance in Fig.~\ref{fig5}(e),
the coreless vortex can still exist in Fig.~\ref{fig6}(e) for $\tilde{a}_{bf}=6$.
The reason is that as the rotation frequency increases, more and more vortices entering in Fig.~\ref{fig6}(e3)
prevent the vortex leaving from the overlapping area.

\begin{figure}[h]
\centering
\includegraphics[width=5in]{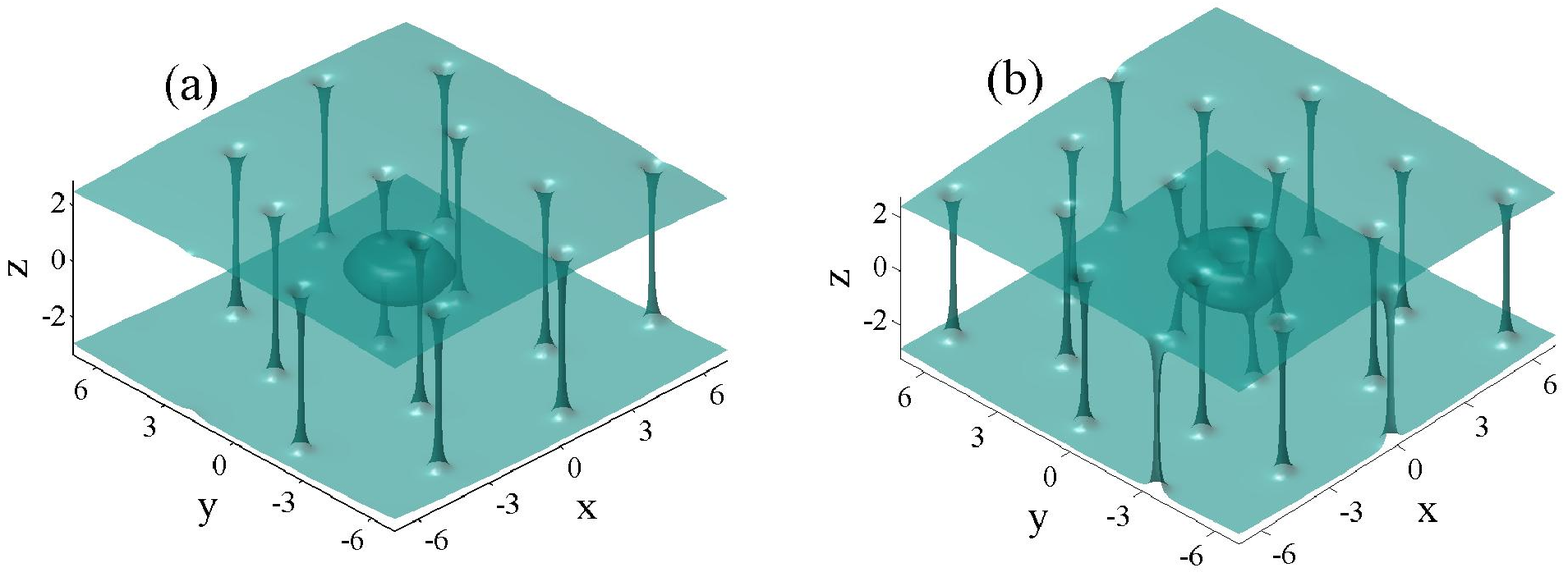}
\caption{\label{fig8} Closeup of 3D visualization (isosurface plot at $1.6\times10^{-4}$)
of the bosonic-fermionic vortices in the phase-separated state. (a) $\Omega=0.35\omega_{f\perp}$ and (b) $0.45\omega_{f\perp}$.}
\end{figure}

In Fig.~\ref{fig7}, we finally study the coupling between the two superfluids containing more vortices for a fast rotation $\Omega=0.45\omega_{f\perp}$.
A very weak boson-fermion interaction ($\tilde{a}_{bf}=0.05$) in Fig.~\ref{fig7}(b) can also lead to
one extra vortex emerging in the Bose superfluid, compared with the single one in Fig.~\ref{fig7}(a).
As the boson-fermion interaction increases in Fig.~\ref{fig7}(c) and \ref{fig7}(d),
similar to Fig.~\ref{fig6}(c) and \ref{fig6}(d),
the Bose superfluids are shrinking with a decrease of the vortex number,
reducing the space of the overlap with the Fermi superfluids and achieving the lower energy.
Such decrease of the vortex number, but a slower one, can be also observed for a mixture with the
number of fermionic pairs decreased to be the same as the bosons (which are not shown here).

In the phase-separated state ($\tilde{a}_{bf}=6$), instead of the coreless vortex in Fig.~\ref{fig6}(e),
the vortex lattice is featured by a new structure as shown in Fig.~\ref{fig7}(e).
It is shown in Fig.~\ref{fig7}(e1) that one of the vortices remains in the Bose superfluid,
which can be verified by the corresponding phase in Fig.~\ref{fig7}(e2). This vortex is surrounded
by three nearest-neighbor vortices in the Fermi superfluid (Fig.~\ref{fig7}(e3)),
which locate at three vertices of the resultant triangular boundary of the Bose superfluid.

Furthermore, the 3D visualization of the new structure is illustrated in Fig.~\ref{fig8}(b).
The vortex lines in the Fermi superfluid pass through the Bose superfluid, bending
towards to the vortex of the Bose superfluid featured by a hole at the center.
It indicates an attractive interaction between the bosonic and fermionic vortices.
The bending of the straight vortex lines in turn result in
the triangular shape of the Bose superfluid.
For comparison, we also plot the 3D visualization of the coreless vortex in Fig.~\ref{fig8}(a),
which is characterized by a straight vortex line of the Fermi superfluid
across the vortex-free Bose superfluid without any deformation.

\section{Conclusions} {\label{secV}}

We have revealed the equilibrium states of the rotating oblate Bose-Fermi superfluid mixtures
in the unitary limit by varying the rotation frequency and the repulsive boson-fermion interaction.
In contrast to the well-known rotating two-component BECs, the $^{41}$K-$^{6}$Li mixture realized experimentally
is a highly asymmetric system. The ratio $0.07/26=0.0027$ between the healing length and radius of the unitary Fermi superfluid
is much smaller than $0.28/3.9=0.08$ of the weakly interacting Bose superfluid.
The critical frequency of the Fermi superfluid $\Omega_c=0.0325\omega_{f\perp}$
is also much smaller than $0.25\omega_{f\perp}$ of the Bose superfluid.
Therefore, the interplay between the bosonic and fermionic vortices with the significant differences
may lead to unique behaviors in quantum instabilities  \cite{raj2023}.

In the very slowly rotating regimes below the critical frequency of
the Bose superfluid, it is clearly that one more vortex can emerge in the Fermi superfluid immediately,
when a very small repulsive boson-fermion interaction is turned on. The vortex number keeps invariant
as the repulsive boson-fermion interaction increases, but the vortex configuration is affected that
the vortices move spirally to the overlapping center area.
Nearby the miscible-immiscible transition one axisymmetic coreless vortex forms, featuring by the vortex core of the Fermi superfluid
filling with the nonrotating Bose superfluid. In the phase-separated state, the coreless vortex
disappears. When entering into the moderate rotating regime,
for a very small repulsive interaction one extra vortex also emerges in the Bose superfluid
as the Fermi superfluid.  As the repulsive boson-fermion interaction further increases,
however, differently from the Fermi superfluids, the Bose superfluids are shrinking with a decrease of the vortex number
to achieve the lower energy. In the phase-separated state
the vortex-lattice structure is finally featured by the coreless vortex.
For the larger rotation frequency, the vortex lattice in the phase-separated state
is instead characterized by a new pattern, which is a single vortex remaining in the Bose superfluid surrounded by three vortices in the Fermi superfluid.
An attractive interaction between the bosonic and fermionic vortices is illustrated through the 3D visualization.
This study not only sheds light on unique phenomena in rotating Bose-Fermi superfluid mixtures, but also provides a theoretical insight into the unconventional behaviors of the vortex numbers arising from the interplay between Bose and Fermi superfluidity observed experimentally.

It should be noted that vortex lattices in strongly interacting Fermi superfluids cannot be resolved {\it in situ},
but detected after ramping the magnetic field to the BEC regime and performing time-flight imaging \cite{sch2007,yao2016}.
The target value of the magnetic field, rate of the ramping, as well as expansion time after the release may profoundly impact
the configurations of vortex lattices in the final state. In a future work it is deserved to study
the effects on vortex-lattice expansions by means of orbital-free
DFT including dissipation \cite{hos2022,cho1998,pro2008}.

\section*{Acknowledgement}
We thank Han Pu for stimulating this work,
and Linhua Wen, Peng Zou and Xiang-Pei Liu for insightful discussions.
This work is supported by the National Natural
Science Foundation of China (NSFC) (Grant No. 12074343, No. 12074120, and No. 11374003)
and Natural Science Foundation of Shanghai (Grant No. 20ZR1418500). The numerical calculations in this paper have
been done on the Tianhe-2 supercomputing system in the national supercomputer center in Guangzhou.


\begin{thebibliography}{00}




\bibitem{pet2008} C. J. Pethick and H. Smith, {\it Bose-Einstein Condensation in Dilute Gases},
2nd ed. (Cambridge University Press, New York, 2008).

\bibitem{fet2001} A. L. Fetter and A. A. Svidzinsky, Vortices in a trapped dilute Bose-Einstein condensate,
J. Phys.: Condens. Matter \textbf{13} (2001) R135.

\bibitem{fet2009} A. L. Fetter, Rotating trapped Bose-Einstein condensates, Rev. Mod. Phys. \textbf{81} (2009) 647.


\bibitem{abo2001} J. R. Abo-Shaeer, C. Raman, J. M. Vogels, and W. Ketterle, Observation of vortex lattices
in Bose-Einstein condensates, Science \textbf{292} (2001) 476.

\bibitem{mad2000} K. W. Madison, F. Chevy, W. Wohlleben, and J. Dalibard, Vortex formation in a stirred Bose-Einstein
condensate, Phys. Rev. Lett. \textbf{84} (2001) 806.


\bibitem{fed2001} D. L. Feder and C. W. Clark, Superfluid-to-solid crossover in a rotating Bose-Einstein condensate,
Phys. Rev. Lett. \textbf{87} (2001) 190401.

\bibitem{mak2003} M. Tsubota, K. Kasamatsu, and M. Ueda, Vortex lattice formation in a rotating Bose-Einstein condensate,
Phys. Rev. A \textbf{65} (2002) 023603.

\bibitem{lobo2004} C. Lobo, A. Sinatra, and Y. Castin, Vortex lattice formation in Bose-Einstein condensates,
Phys. Rev. Lett. \textbf{92} (2004) 020403.


\bibitem{blo2009} I. Bloch, J. Dalibard, and W. Zwerger, Many-body physics with ultracold gases, Rev. Mod. Phys.
\textbf{80} (2008) 885.

\bibitem{zai2009} H. Zhai, Strongly interacting ultracold quantum gases, Front. Phys. China \textbf{4} (2009) 1.

\bibitem{zwe2012} W. Zwerger (Ed.), {\it The BCS-BEC Crossover and the Unitary Fermi Gas (Lecture Notes in Physics vol 836)}, (Springer, Berlin, 2012).


\bibitem{raj2023} D. H.-Rajkov, N. Grani, F. Scazza, G. D. Pace, W. J. Kwon, M. Inguscio, K. Xhani, C. Fort, M. Modugno, F. Marino,
and G. Roati, Universality of the superfliud Kelvin-Helmholtz instability by single-vortex tracking,  arXiv: 2303.12631v1.

\bibitem{pac2022} G. D. Pace, K. Xhani, A. M. Falconi, M. Fedrizzi, N. Grani, D. H. Rajkov, M. Inguscio, F. Scazza, W. J. Kwon, and
G. Roati, Imprinting persistent currents in tunable fermionic rings, Phys. Rev. X \textbf{12} (2022) 041037.


\bibitem{yan2022} Y. Cai, D. G. Allman, P. Sabharwal, and K. C. Wright, Persistent currents in rings of ultracold fermionic atoms,
Phys. Rev. Lett. \textbf{128} (2022) 150401.

\bibitem{xiang2021} X.-P. Liu, X.-C. Yao, Y. Deng, Y.-X. Wang, X.-Q. Wang, X. Li, Q. Chen, Y.-A, Chen, and J.-W. Pan,
Dynamic formation of quasicondensate and spontaneous vortices in a strongly interacting Fermi gas, Phys.
Rev. Research \textbf{3} (2022) 043115.

\bibitem{fang2021} F. Li, S. Deng, L. Zhang, J. Xia, L. Yi, and H. Wu, Light induced space-time patterns in a superfluid Fermi gas,
Sci. China-Phys. Mech. Astron. \textbf{64} (2021) 294212.

\bibitem{die2021} D. H.-Rajkov, J. E. P.-Castillo, A. d. R.-Lima, A. G.-Vald\'{e}s, F. J. P.-Cuevas and J. A. Seman,
Faraday waves in strongly interacting superfluids, New J. Phys. {\bf 23} (2021) 103038.


\bibitem{zwi2005} M. W. Zwierlein, J. R. A.-Shaeer, A. Schirotzek, C. H. Schunck, and W. Ketterle, Vortices and superfluidity in a strongly
interacting Fermi gas, Nature  \textbf{435} (2021) 1047.

\bibitem{sch2007} C. H. Schunck, M. W. Zwierlein, A. Schirotzek, and W. Ketterle, Superfluid expansion of a rotating Fermi gas,
Phys. Rev. Lett. \textbf{98} (2007) 050404.



\bibitem{gio2008} S. Giorgini, L. P. Pitaevskii and S. Stringari, Theory of ultracold atomic Fermi gases, Rev. Mod. Phys. \textbf{80} (2008) 1215.

\bibitem{abul2012} A. Bulgac, M. M. Forbes, and P. Magierski, The unitary Fermi gas: From Monte Carlo to density
functionals, in {\it The BCS-BEC Crossover and the Unitary Fermi Gas}, edited by W. Zwerger (Springer, Berlin, 2012).

\bibitem{dav2004} D. L. Feder, Vortex arrays in a rotating superfluid Fermi gas, Phys. Rev. Lett. \textbf{93} (2004) 200406.

\bibitem{sim2015} S. Simonucci, P. Pieri, and G. C. Strinati, Vortex arrays in neutral trapped Fermi gases through the BCS-BEC crossover,
Nat. Phys. \textbf{11} (2015) 941.

\bibitem{hu2007} H. Hu and X.-J. Liu, Density fingerprint of giant vortices in Fermi gases near a Feshbach resonance,
Phys. Rev. A \textbf{75} (2007) 011603(R).

\bibitem{kon2021} L. Kong, G. Fan, S.-G. Peng, X.-L. Chen, H. Zhao, and P. Zou, Dynamical generation of solitons in one-dimensional
Fermi superfluids with and without spin-orbit coupling, Phys. Rev. A \textbf{103} (2007) 063318.


\bibitem{fan2022} G. Fan. X.-L. Chen and P. Zou, Probing two Higgs oscillations in a one-dimensional Fermi superfluid with Raman-type
spin-orbit coupling, Front. Phys. \textbf{17} (2022) 52502.

\bibitem{aur2003} A. Bulgac and Y. Yu, Vortex state in a strongly coupled dilute atomic fermionic superfluid,
Phys. Rev. Lett. \textbf{91} (2003) 190404.

\bibitem{abu2011} A. Bulgac, Y.-L. Luo, P. Magierski, K. J. Roche, and Y. Yu, Real-time dynamics of quantized
vortices in a unitary Fermi superfluid, Science {\bf 332} (2011) 1288.

\bibitem{bul2014} A. Bulgac, M. M. Forbes, M. M. Kelley, K. J. Roche, and G. Wlaz{\l}owski, Quantized superfluid
vortex rings in the unitary Fermi gas, Phys. Rev. Lett. \textbf{112} (2014) 025301.

\bibitem{hos2022} K. Hossain, K. Kobuszewski, M. M. Forbes, P. Magierski, K. Sekizawa, and G. Wlaz{\l}owski,
Rotating quantum turbulence in the unitary Fermi gas, Phys. Rev. A \textbf{105} (2022) 013304.

\bibitem{kop2021} J. Kopyci\'{n}ski, W. R. Pudelko, and G. Wlaz{\l}owski, Vortex lattice in spin-imbalanced unitary Fermi gas,
Phys. Rev. A \textbf{104} (2021) 053322.

\bibitem{bar2023} A. Barresi, A. Boulet, P. Magierski, and G. Wlaz{\l}owski, Dissipative dynamics of quantum vortices in fermionic superfluid,
Phys. Rev. Lett. \textbf{130} (2021) 043001.


\bibitem{fer2014} I. F.-Barbut, M. Delehaye, S. Laurent, A. T. Grier, M. Pierce, B. S. Rem, F. Chevy, and C. Salomon,
A mixture of Bose and Fermi superfluids, Science \textbf{345} (2014) 1035.

\bibitem{tak2016} T. Ikemachi, A. Ito, Y. Aratake, Y. Chen, M. Koashi, M. K.-Gonokami and M. Horikoshi,
All-optical production of dual Bose-Einstein condensates of paired fermions and bosons with $^6$Li and $^7$Li,
J. Phys. B: At. Mol. Opt. Phys. {\bf 50} (2014) 01LT01.

\bibitem{yao2016} X.-C. Yao, H.-Z. Chen, Y.-P. Wu, X.-P. Liu, X.-Q. Wang, X. Jiang, Y. Deng, Y.-A. Chen, and J.-W. Pan,
Observation of coupled vortex lattices in a mass-imbalance Bose and Fermi superfluid mixture,
Phys. Rev. Lett. \textbf{117} (2014) 145301.

\bibitem{roy2017} R. Roy, A. Green, R. Bowler, and S. Gupta, Two-element mixture of Bose and Fermi superfluids,
Phys. Rev. Lett. \textbf{118} (2014) 055301.


\bibitem{mat1999} M. R. Matthews, B. P. Anderson, P. C. Haljan, D. S. Hall, C. E. Wieman, and E. A. Cornell,
Vortices in a Bose-Einstein condensate, Phys. Rev. Lett. \textbf{83} (1999) 2498.

\bibitem{and2000} B. P. Anderson, P. C. Haljan, C. E. Wieman, and E. A. Cornell, Vortex precession in Bose-Einstein condensates:
obervations with filled and empty cores, Phys. Rev. Lett. \textbf{85} (2000) 2857.

\bibitem{sch2004} V. Schweikhard, I. Coddington, P. Engels, S. Tung, and E. A. Cornell, Vortex-lattice dynamics in rotating spinor
Bose-Einstein condensates, Phys. Rev. Lett. \textbf{93}  (2004) 210403.


\bibitem{kas2005} K. Kasamatsu, M. Tsubota, and M. Ueda, Vortices in multicomponent Bose-Einstein condensates,
Int. J. Mod. Phys. B \textbf{19} (2005) 1835.

\bibitem{pu1997} H. Pu and N. P. Bigelow, Properties of two-species Bose condensates, Phys. Rev. Lett. \textbf{80} (1998) 1130.

\bibitem{muel2002} E. J. Mueller and T. L. Ho, Two-component Bose-Einstein condensates with
a large number of vortices, Phys. Rev. Lett. \textbf{88} (2002) 180403.

\bibitem{kas2003} K. Kasamatsu, M. Tsubota, and M. Ueda, Vortex phase diagram in rotating two-component
Bose-Einstein condensate, Phys. Rev. Lett. \textbf{91} (2002) 150406.

\bibitem{kas2009} K. Kasamatsu and M. Tsubota, Votex sheet in rotating two-component Bose-Einstein condensates,
Phys. Rev. A \textbf{79} (2009) 023606.

\bibitem{ken2018} K. Kasamatsu and K. Sakashita, Stripes and honeycomb lattice of quantized vortices in rotating two-component
Bose-Einstein condensates, Phys. Rev. A \textbf{97} (2018) 053622.

\bibitem{mas2011} P. Mason and A. Aftalion, Classification of the ground states and topological defects in
a rotating two-component Bose-Einstein condensate, Phys. Rev. A \textbf{84} (2011) 033611.

\bibitem{aft2012} A. Aftalion, P. Mason, and J. Wei, Vortex-peak interaction and lattice shape
in rotating two-component Bose-Einstein condensates, Phys. Rev. A \textbf{85} (2012) 033614.

\bibitem{ska2020} S. K. Adhikari, Phase-separated symmetry-breaking vortex-lattice in a binary Bose-Einstein condensate,
Physica E \textbf{115} (2020) 113713.

\bibitem{min2019} L. Mingarelli and R. Barnett, Exotic vortex lattices in binary repulsive superfluids,
Phys. Rev. Lett. \textbf{122} (2019) 045301.

\bibitem{min2018} L. Mingarelli, E. E. Keaveny and R. Barnett, Vortex lattices in binary mixtures of
repulsive superfluids, Phys. Rev. A \textbf{97} (2018) 043622.



\bibitem{ling2014} L. Wen and J. Li, Structure of dynamics of a rotating superfluid Bose-Fermi mixture,
Phys. Rev. A \textbf{90} (2014) 053621.

\bibitem{jiang2017} Y. Jiang, R. Qi, Z.-Y. Shi, and H. Zhai, Vortex lattices in the Bose-Fermi superfluid mixture,
Phys. Rev. Lett. \textbf{118} (2017) 080403.

\bibitem{pan2017} J.-S. Pan, W. Zhang, W. Yi and G.-C. Guo, Vortex-core structure in a mixture of Bose and Fermi superfluids,
Phys. Rev. A \textbf{95} (2017) 063614.

\bibitem{ogr2020} M. \"{O}gren and G. M. Kavoulakis, Rotational properties of superfluid Fermi-Bose mixtures in a tight torodial trap,
Phys. Rev. A \textbf{102} (2017) 013323.


\bibitem{kim2004} Y. E. Kim and A. L. Zubarev, Time-dependent density-functional theory for trapped strongly interacting fermionic atoms, Phys. Rev. A \textbf{70} (2017) 033612.

\bibitem{sal2008} L. Salasnich, N. Manini, and F. Toigo, Macroscopic periodic tunneling of Fermi atoms in the BCS-BEC crossover,
Phys. Rev. A \textbf{77} (2008) 043609.

\bibitem{lsal2008} L. Salasnich and F. Toigo, Extended Thomas-Fermi density functional for the unitary Fermi gas,
Phys. Rev. A \textbf{78} (2008) 053626.

\bibitem{adh2008} S. K. Adhikari, Nonlinear Schr\"{o}dinger equation for a superfluid Fermi gas in the BCS-BEC crossover,
Phys. Rev. A \textbf{77} (2008) 045602.

\bibitem{wen2008} W. Wen, Y. Zhou, and G. Huang, Interference patterns of superfluid Fermi gases in the BCS-BEC crossover
released from optical lattices, Phys. Rev. A \textbf{77} (2008) 033623.

\bibitem{sadh2008} S. K. Adhikari and L. Salasnich, Superfluid Bose-Fermi mixture from weak coupling to unitarity,
Phys. Rev. A \textbf{78} (2008) 043616.

\bibitem{adh2010} S. K. Adhikari, B. A. Malomed, L. Salasnich, and F. Toigo, Spontaneous symmetry breaking of Bose-Fermi mixtures
in double-well potentials, Phys. Rev. A \textbf{81} (2010) 053630.

\bibitem{yon2011} Y. Cheng and S. K. Adhikari, Localization of a Bose-Fermi mixture in a bichromatic optical lattice,
Phys. Rev. A \textbf{84} (2011) 023632.

\bibitem{wen2018} W. Wen and H.-j. Li, Collective dipole oscillations in a mixture of Bose and Fermi
superfluids in the BCS-BEC crossover, New J. Phys. \textbf{20} (2018) 083044.

\bibitem{kho2021} K. Hossain, S. Gupta, and M. M. Forbes, Detecting entrainment in Fermi-Bose mixtures,
Phys. Rev. A \textbf{105} (2022) 063315.


\bibitem{leg2006} A. J. Leggett, {\it Quantum Liquids: Bose Condensation and Cooper Pairing in Condensed Matter Systems} (
Oxford University Press, Oxford, 2006).


\bibitem{mmf2014} M. M. Forbes and R. Sharma, Validating simple dynamical simulations of the unitary Fermi gas,
Phys. Rev. A \textbf{90} (2014) 043638.


\bibitem{wen2013}W. Wen, C. Zhao and X. Ma, Dark-soliton dynamics and snake instablity in superfluid Fermi gaes
trapped by an anisotropic harmonic potential, Phys. Rev. A \textbf{88} (2013) 063621.


\bibitem{nav2010} N. Navon, S. Nascimb\`{e}ne, F. Chevy, and C. Salomon, The equation of state of a low-temperature
Fermi gas with tunable interactions, Science \textbf{328} (2010) 729.

\bibitem{mani2005} N. Manini and L. Salasnich, Bulk and collective properties of a dilute Fermi gas in the BCS-BEC
crossover, Phys. Rev. A \textbf{71} (2005) 033625.

\bibitem{wen2010} W. Wen, S.-Q. Shen, and G. Huang, Propagation of sound and supersonic bright solitons in superfluid Fermi gases in BCS-BEC crossover, Phys. Rev. B \textbf{81} (2010) 014528.


\bibitem{ram2019} R. K. Kumar, V. Lon\v{c}ar, P. Muruganandam, S. K. Adhikari, and A. Bala\v{z}, C and Fortran
OpenMP programs for rotating Bose-Einstein condensates, Comput. Phys. Commun. \textbf{240} (2019) 74.

\bibitem{kum2017} R. K. Kumar, L. Tomio, B. A. Malomed, and A. Gammal, Vortex lattices in binary Bose-Einstein condensates with dipole-dipole interactions, Phys. Rev. A \textbf{96} (2017) 063624.

\bibitem{mur2009} P. Muruganandam and S. K. Adhikari, Fortran programs for the time-dependent Gross-Pitaevskii
equation in a fully anisotropic trap,  Comput. Phys. Commun. \textbf{180} (2009) 1888.

\bibitem{sat2015} B. Satari\'{a}, V. Slavni\'{c}, A. Beli\'{c}, A. Bala\v{z}, P. Muruganandam, and S. K. Adhikari,
Hybrid OpenMP/MPI programs for solving the time-dependent Gross-Pitaevskii equation in a fully anisotropic trap,
Comput. Phys. Commun. \textbf{200} (2016) 411.


\bibitem{fey1955} R. P. Feynman, in {\it Progress in Low Temperature Physics}, edited by C. J. Gorter (North-Holland, Amsterdam, 1955),
Chap. 2.


\bibitem{zhang2020} L. Zhang, W. Wen, J. Qian, X. Ma and Y. Wang, Anisotropic expansions of a strongly interacting
Fermi superfluid containing a vortex, J. Phys. B: At. Mol. Opt. Phys. \textbf{53} (2020) 155304.




\bibitem{bru2001} G. M. Brunn and L. Viverit, Vortex state in superfluid trapped Fermi gases at zero temperature,
Phys. Rev. A \textbf{64} (2001) 063606.


\bibitem{hui2006} H. Zhai and T.-L. Ho, Critical rotational frequency for superfluid fermionic gases across a Feshbach resonance,
Phys. Rev. Lett. \textbf{97} (2001) 180414.


\bibitem{bho2008} S. G. Bhongale and H. Pu, Phase separation in a mixture of a Bose-Einstein condensate and a
two-component Fermi gas as a probe of Fermi superfluidity, Phys. Rev. A \textbf{78} (2008) 061606(R).

\bibitem{fla2007} L. Salasnich and F. Toigo, Fermi-Bose mixture across a Feshbach resonance, Phys. Rev. A \textbf{75} (2007) 013623.


\bibitem{keni2005} K. Kasamatsu, M. Tsubota, and M. Ueda, Spin textures in rotating two-component Bose-Einstein condensates, Phys. Rev. A  \textbf{71} (2005) 043611.

\bibitem{Ric2020} A. Richaud, V. Penna, R. Mayol, and M. Guilleumas, Vortices with massive cores in a binary mixture of Bose-Einstein condensates, Phys. Rev. A  \textbf{101} (2020) 013630.

\bibitem{Ric2021} A. Richaud, V. Penna, and A. L. Fetter, Dynamics of massive point vortices in a binary mixture of Bose-Einstein condensates, Phys. Rev. A  \textbf{103} (2021) 023311.

\bibitem{cho1998} S. Choi, S. A. Morgan, and K. Burnett, Phenomenological damping in trapped atomic Bose-Einstein
condensates, Phys. Rev. A \textbf{57} (1998) 4057.

\bibitem{pro2008} N. P. Proukakis and B. Jackson, Finite-temperature models of Bose-Einstein condensation, J. Phys.
B: At. Mol. Opt. Phys. \textbf{41} (2008) 203002.



\end{thebibliography}
\end{document}